\documentclass[conference]{IEEEtran}

\usepackage{amsmath,amssymb,amsfonts}
\usepackage[linesnumbered,ruled,vlined]{algorithm2e}
\usepackage{graphicx}
\usepackage{textcomp}
\usepackage{listings}
\usepackage{fancyhdr}
\usepackage{comment}
\usepackage{mathptmx} 
\usepackage{soul}
\usepackage{multirow}
\usepackage{booktabs}
\usepackage[normalem]{ulem}
\usepackage[sort,nocompress]{cite}
\usepackage[T1]{fontenc}
\usepackage{amssymb}
\usepackage{pifont}
\usepackage{tablefootnote}
\usepackage[T1]{fontenc}

\usepackage[hyphens]{url}
\usepackage[final]{microtype}
\usepackage[table,xcdraw,dvipsnames]{xcolor}
\usepackage[keeplastbox]{flushend}

\usepackage{enumitem}
\setlist[enumerate]{noitemsep}

\usepackage[bookmarks=true,breaklinks=true,letterpaper=true,colorlinks,linkcolor=blue,citecolor=blue,urlcolor=blue]{hyperref}

\newcommand{\insertFigure}[2]{
    \begin{figure}[t!]
        \centering
        \includegraphics[width=\linewidth]{figures/#1.pdf}
	\vspace{-6mm}
        \caption{\small #2}
	\vspace{-3mm}
        \label{fig:#1}
    \end{figure}
}

\newcommand{\insertWideFigure}[2]{

    \begin{figure*}[ht!]
        \centering
        \includegraphics[width=\textwidth]{figures/#1.pdf}
	\vspace{-6mm}
        \caption{\small #2}
	\vspace{-3mm}
        \label{fig:#1}
    \end{figure*}

}

\newcommand{\squishlist}{
 \begin{list}{$\bullet$}
  { \setlength{\itemsep}{0pt}
     \setlength{\parsep}{0pt}
     \setlength{\topsep}{3pt}
     \setlength{\partopsep}{0pt}
     \setlength{\leftmargin}{1.5em}
     \setlength{\labelwidth}{1em}
     \setlength{\labelsep}{0.5em} } }
     
\newcommand{\squishnums}{
 \begin{list}{$\bullets$}
  { \setlength{\itemsep}{0pt}
     \setlength{\parsep}{3pt}
     \setlength{\topsep}{3pt}
     \setlength{\partopsep}{0pt}
     \setlength{\leftmargin}{1.5em}
     \setlength{\labelwidth}{1em}
     \setlength{\labelsep}{0.5em} } }

\newcommand{\squishlisttwo}{
 \begin{list}{$\bullet$}
  { \setlength{\itemsep}{0pt}
     \setlength{\parsep}{0pt}
    \setlength{\topsep}{0pt}
    \setlength{\partopsep}{0pt}
    \setlength{\leftmargin}{2em}
    \setlength{\labelwidth}{1.5em}
    \setlength{\labelsep}{0.5em} } }

\newcommand{\squishend}{
  \end{list}  }

\newcommand{\TitleNamenospace}[0]{\textsc{Harp}}

\newcommand{\HHPName}[0]{\textsc{Harp}~}

\newcommand{\HHPNamenospace}[0]{\textsc{Harp}}

\newcommand{\TODO}[1]{\textcolor{red}{TODO::: #1}}
\newcommand{\RG}[1]{{\color{orange}\bfseries [RG::: #1]}}
\newcommand{\TK}[1]{{\color{teal}\bfseries [Tushar::: #1]}}

\renewcommand{\RG}[1]{\ignorespaces}

\renewcommand{\TK}[1]{\ignorespaces}

\renewcommand{\TODO}[1]{\ignorespaces}

\newcommand{\insertWideFigureScaled}[3]{

    \begin{figure*}[ht!]
        \centering
        \includegraphics[width=#3\textwidth]{figures/#1.pdf}
	\vspace{-3mm}
        \caption{\small #2}
	\vspace{0mm}
        \label{fig:#1}
    \end{figure*}

}

\newcommand{\insertFigureScaled}[3]{

    \begin{figure}[t!]
        \centering
        \includegraphics[width=#3\linewidth]{figures/#1.pdf}
	\vspace{-3mm}
        \caption{\small #2}
	\vspace{0mm}
        \label{fig:#1}
    \end{figure}

}
\usepackage{url}

\newcommand{\microsubmissionnumber}{xxx} 

\fancypagestyle{firstpage}{
  \fancyhf{}
  
  \fancyhead[C]{\vspace{10pt}\normalsize{ISCA 2025 Submission
      \textbf{\#\microsubmissionnumber} -- Confidential Draft -- Do NOT Distribute!!}\\\vspace{-25pt}} 
  \fancyfoot[C]{\thepage}
}

\title{\TitleNamenospace: A Taxonomy for \underline{H}eterogeneous and Hier\underline{ar}chical \underline{P}rocessors for Mixed-reuse Workloads
}

\author{\IEEEauthorblockN{Raveesh Garg}
\IEEEauthorblockA{
\textit{Georgia Institute of Technology}\\
raveesh.g@gatech.edu}
\and
\IEEEauthorblockN{Michael Pellauer}
\IEEEauthorblockA{
\textit{NVIDIA}\\
mpellauer@nvidia.com}
\and
\IEEEauthorblockN{Tushar Krishna}
\IEEEauthorblockA{
\textit{Georgia Institute of Technology}\\
tushar@ece.gatech.edu}
}

\begin{document}
\date{}

\maketitle
\pagestyle{plain}
\begin{abstract}
    Artificial intelligence (AI) application domains consist of a mix of tensor operations with high and low arithmetic intensities (aka reuse). Hierarchical (i.e. compute along multiple levels of memory hierarchy) and heterogeneous (multiple different sub-accelerators) accelerators are emerging as a popular way to process mixed reuse workloads, and workloads which consist of tensor operators with diverse shapes. 
    However, the space of hierarchical and/or heterogeneous processors (HHP's) is relatively under-explored.  Prior works have proposed custom architectures to take advantage of heterogeneity to have multiple sub-accelerators that are efficient for different operator shapes. In this work, we propose \HHPNamenospace, a taxonomy to classify various hierarchical and heterogeneous accelerators and use the it to study the impact of heterogeneity at various levels in the architecture. \HHPName taxonomy captures various ways in which HHP's can be conceived, ranging from B100 cores with an "intra-node heterogeneity" between SM and tensor core to NeuPIM with cross-depth heterogeneity which occurs at different levels of memory hierarchy. We use Timeloop mapper to find the best mapping for sub-accelerators and also modify the timeloop cost model to extend it to model hierarchical and heterogeneous accelerators.

\end{abstract}

\section{Introduction}
\label{sec:introduction}

Artificial intelligence (AI) applications consist of  cascades of tensor operations of various shapes and sizes~\cite{kwon2018maeri,kwon2019understanding,flat}. As a result, these cascades have varying arithmetic intensities, i.e. reuse. An obvious example of this is transformers~\cite{attention} with multi-head attention using batched matrix multiplication (BMM) and layer normalization which have lower reuse than GEMMs involved in Q,K,V generation and feedforward layers. Autoregressive decoding~\cite{gpt4} has a decode phase which has less reuse than BMMs in the summarization phase~\cite{neupim}. In domains such as AR/VR, operations have differences in the range of arithmetic intensities, and even individual AI models (task) have different overall reuse trends~\cite{xrbench}. Recent applications like neurosymbolic AI~\cite{neurosymbolic} combine probabilistic models and DNN models. Therefore, AI workloads exhibit mixed reuse (i.e., a mix of high and low reuse). 

Earlier accelerators~\cite{kwon2018maeri,tpu-isca,eyeriss2016isca,nvdla} and mapping frameworks~\cite{kao2020gamma,cosa,kwon2019understanding,interstellar,timeloop,pipeorgan} were able to exploit reuse on earlier CNN layers with cubic aspect ratios. However, with the emergence of DNN application domains such as AR / VR~\cite{xrbench} and applications such as GNNs~\cite{hamilton2017inductive} and transformers, applications have operations with low arithmetic intensity and hence prior works have used inter-operation fusion/pipelining~\cite{garg2021understanding,flat,flashattention,pipeorgan,tileflow,isca-pip,isos,tangram} which involves reusing the intermediate tiles by staging them on-chip. However, the arithmetic intensities of even a fused cascade of a few operations are low. For example, in the decode stage of autoregressive decoding, the arithmetic intensity of the operators is 1-2 orders of magnitude lower than the arithmetic intensity required to saturate the datapath~\cite{neupim}. These works also miss the opportunity of hiding low-reuse operations behind high-reuse operations to keep datapath utilization high.

To this end, hierarchical heterogeneous processors (HHP's, term coined by Symphony~\cite{symphony}) have emerged as an attractive solution, particularly for mixed-intensity workloads, where low-intensity operators can be hidden behind high-intensity operators. Recent works NeuPIM~\cite{neupim} and Duplex~\cite{duplex} proposed heterogeneous accelerators with different sub-accelerators at different levels of memory hierarchy. Symphony~\cite{symphony} proposed another hierarchical heterogeneous accelerator for sparse and dense processing on one accelerator substrate.
Mapping different operations on different sub-accelerators decouples low-reuse operations from the critical path by overlapping it with high-reuse operations, thus mitigating its impact. 
Thus, HHP's can concurrently running high-reuse and low-reuse operations uses resources in a complementary fashion. High-reuse operations require low memory bandwidth but high on-chip memory to reuse. Low-reuse operations require a higher memory bandwidth and a lower on-chip memory, just enough to hide the latency.~\autoref{fig:roofline} shows the compute roof and bandwidth partitioning example on a roofline.

With different HHP's being proposed, we aim to classify the HHP's under a common taxonomy, and explore the impact of architectural choices on performance and energy efficiency. 


Prior works have heterogeneous architectures with vastly different hardware organizations.
 The most popular example of heterogeneous architecture is the NVIDIA tensor core chip~\cite{blackwell} where there is heterogeneity between streaming multiprocessors and tensor cores. Moreover, this is the most tightly coupled form of heterogeneity, since both tensor core and streaming multiprocessors are governed by a single FSM controller. NeuPIM~\cite{neupim} and Duplex~\cite{duplex} are at the other end of the spectrum where the heterogeneous sub-accelerators are at different levels of memory hierarchy. This essentially means that for a system with NPU and in-DRAM/near-DRAM processor, the only shared resource is the DRAM, while for a near-LLB (Last Level Buffer) accelerator, the only shared resources are last-level buffers and DRAM. Another major difference between these architectures is that NVIDIA tensor-cores have compute only at the leaves of the memory hierarchy, while architectures like NeuPIM have compute across memory hierarchy.

Differences in these architectures have an impact on the performance of different workloads. For example, NeuPIM-like accelerators are more desirable for workloads with clearly marked high- and low-reuse operations running concurrently on separate sub-accelerators (e.g. prefill and decode phases of decoder-only transformers for workloads like GPT3~\cite{gpt3} and Llama~\cite{llama}). On another extreme, for traditional DNN's with sufficiently cubic aspect ratios, a homogeneous accelerator with compute only at the leaves, provides the highest undivided throughput. Moreover, with sufficiently high memory bandwidth, a homogeneous accelerator slightly edges a heterogeneous accelerator for an encoder-only mixed reuse transformer (e.g. BERT~\cite{bert}) where the arithmetic intensity differences gap between high- and low-reuse operations is less than an order of magnitude, and the dependencies make it hard to achieve perfect decoupling of high-reuse and low-reuse operations. Another factor that influences these architectures is mapping, particularly mapping constraints, coupled with the microarchitecture. For example, a heterogeneous accelerator could be sensitive to the mapping constraints in low-reuse sub-accelerator, but less sensitive to mapping constraints in the high-reuse sub-accelerator (because of sufficient dimension sizes), which is the case with decoder-only transformers.


In this work, \textit{we propose a \HHPNamenospace\footnote{\underline{H}eterogeneous and Hier\underline{AR}chical \underline{P}rocessors}, a taxonomy to classify the heterogeneous and hierarchical  accelerators.}
This provides a systematic approach to characterize hardware organization of different hierarchical and/or heterogeneous accelerators.
Specifically, the taxonomy classifies the accelerators based on relative location of sub-accelerators w.r.t. their FSM and memory hierarchy and on the basis of whether the compute is only at the leaf or throughout memory hierarchy.
We can also use the taxonomy to derive a new class of accelerators, which prior work has not exhibited.

We modify Timeloop~\cite{timeloop} and develop a wrapper around it to model the performance for various hierarchical and/or heterogeneous accelerator configurations described by our taxonomy. We compare several different hierarchical and/or heterogeneous architecture design points for various mixed-reuse tensor workloads.
We also explore the detailed impact of \textit{partitioning resources (e.g. memory bandwidth) across sub-accelerators} in a configuration.

\section{Background}
\label{sec:background}

\subsection{Mapping}

Mapping refers to the a set of loop-transformations to schedule the workload on spatial accelerator. Mapping comprises of different aspects like loop permutation, parallelization strategies, tiling strategies (slicing a dimension into multiple loop nests depending on buffer sizes) etc. Mapping strategies mainly affect the static utilization of the PE-array, and the reuse achieved.

\subsection{Mixed-reuse Workloads}

Mixed-reuse workloads refer to workloads which have both high arithmetic intensity and low arithmetic intensity operations. Unlike earlier DNNs which have high- intensity operations or scientific applications like CG which have low- intensity operations, mixed-reuse workloads have both high- and low-intensity operations. One of the examples is encoder-only transformer (e.g. BERT) where Q/K/V generation and deprojection layers are high arithmetic intensity operations while logit and attend have relatively low arithmetic intensity. However, in BERT, there are sequential dependencies between high- and low-reuse operations, and the only operations that can overlap are logit ($P=QK^T$) and value generation operations ($V=IW_v$).

 \insertFigure{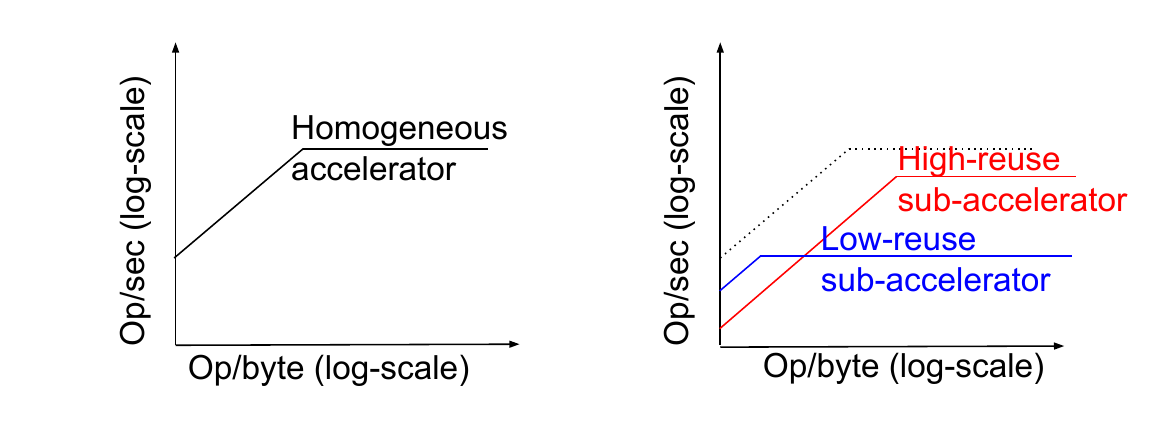}{Roofline in heterogeneous accelerator with high- and low-reuse sub-acceelrators compared to homogeneous accelerator with total area and memory bandwidth.}

Decoder-only transformers (e.g. GPT-3, Llama), on the other hand, have pre-fill and decode stages, and can be pipelined at granularity of single batch. Prefill stage in terms of einsums, is similar to the encoder-only transformer with Q/K/V generation, logit, attend, deprojection and feed-forward network. In the decode stage, new tokens are generated for each query one at a time. As a result, the sequence length on query side is 1, and the process is repeated until all tokens are generated. As a result, decode stage consists of multiple tensor operations with smaller aspect ratios with lower MACs as well as proportionally lower arithmetic intensity. These attention layers are repeated serially for each token. Since prefill and decode stages for a group of tokens can be decoupled in a coarse grained manner, mapping prefill stage on a high-reuse accelerator and decode stage on a low-reuse accelerator provides much better decoupling of high- and low-reuse operations than encoder-only models. Throughout this work, we discuss the impact of this decoupling on homogeneous, heterogeneous and hierarchical accelerators.

\section{Insights and Motivation}
\label{sec:motivation}

\subsection{Impact of Paritioning across Sub-accelerators}

As discussed, one of the ways of partitioning the workload on HHP sub-accelerators is based on reuse, with a high-reuse accelerator running high reuse component decoupled from the low reuse accelerator running low reuse component. The case for partitioning the workload by reuse is that high-reuse and low-reuse operations use resources in a complementary fashion.

\subsubsection{Impact on roofline}

 High-reuse operations in the application are highly compute-bound and end up saturating the bandwidth even with a fraction of peak memory bandwidth. $BW_{high-reuse}=BW_{peak}\times \frac{AI_{tipping}}{AI_{op}}$. On the other hand, low-reuse operations are bounded by memory bandwidth and performance in the memory-bound region is proportional to the memory bandwidth, and hence these operations are more likely to use the memory bandwidth.~\autoref{fig:roofline} shows how the roofline is split in heterogeneous architectures with a high-reuse and low-reuse sub-accelerator compared to a plain homogeneous accelerator with same total compute roof. High-reuse sub-accelerator has a higher compute roof and requires a lower memory bandwidth, as a result of which it can be compute-bound even if the tipping point were higher. For a low-reuse accelerator, the performance is directly proportional to memory bandwidth. So in order to minimize major dark silicon in situations where high-reuse sub-accelerator is idle waiting for low-reuse accelerator to finish, a low-reuse accelerator is granted a higher portion of the memory bandwidth. One of the major caveats with this kind of split though is the overlap opportunity. It is possible that only 20\% of the whole cascade of operations can be run on high- and low-reuse sub-accelerators independently. In this case, the underutilization due to dark silicon is significant. However, if cascade can be decoupled into larger independent sub-cascades, mapping high- and low-reuse operations on a heterogeneous accelerator is beneficial compared to a homogeneous accelerator, since the latter gets massively underutilized during a low-reuse sub-cascade of operations.


 \insertFigureScaled{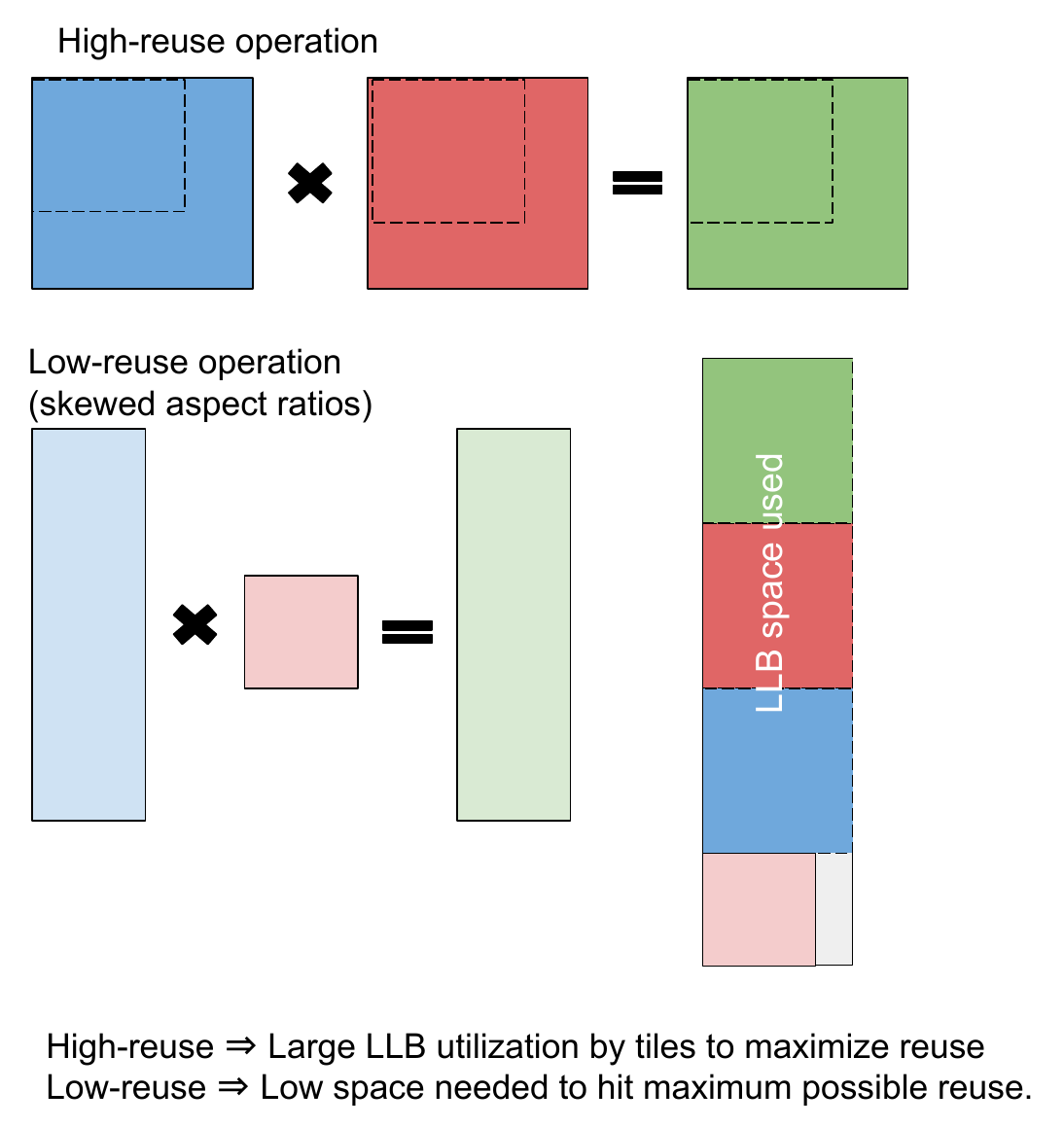}{Partitioning of LLB resources.}{0.9}

 \subsubsection{Impact on LLB utilization}
 
LLB usage of these operations trends in the opposite direction to roofline. High-reuse operations need vast LLB space in order to make use of the reuse opportunity due to the large dimensions of the matrix, while low reuse operations easily hit the peak arithmetic intensity when the smallest tensor fits the LLB requiring a fraction of LLB space, as~\autoref{fig:resources}(b) shows. Therefore, high-reuse and low-reuse partitioning is a synergetic way to allocate operations.

\insertWideFigureScaled{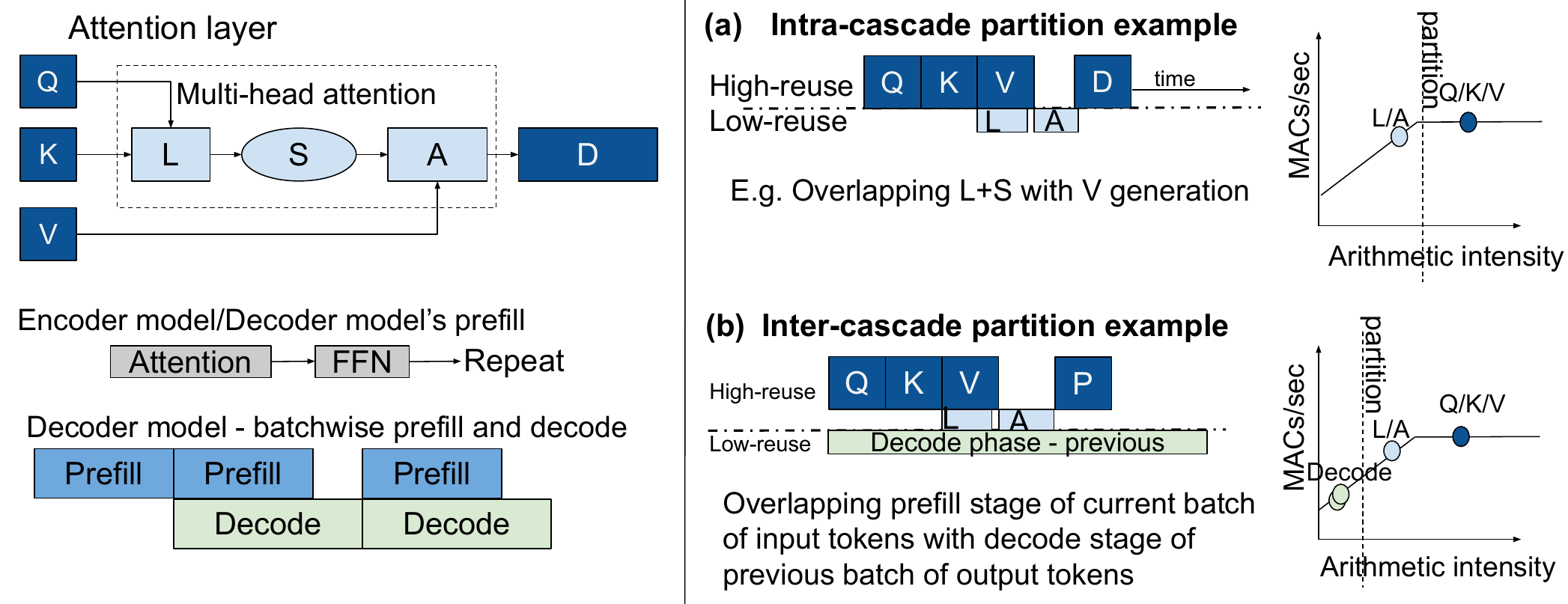}{(a) Intra-cascade partitioning within the encoder model. (b) Inter-cascade partitioning of the decoder model into prefill and decode phases with the prefill phase mapped on high-reuse sub-accelerator and decode phase mapped on low-reuse sub-accelerator.}{.95}

\subsection{Examples of Workload Partitioning}

\textbf{Intra-Cascade Partitioning:}
\autoref{fig:partitioning}(a) shows example workload partitioning strategies in high- and low-reuse operations. High-reuse sub-accelerator runs the high-reuse operations (e.g. GEMMs in encoder models and prefill phase of decoder models etc.) while low-reuse sub-accelerator runs the low-reuse operations (e.g. Multi-head attention, entire decode phase of decoder models). As far as load distribution is concerned, since the high-reuse sub-accelerator has a larger number of PEs, prioritizing keeping the high-reuse sub-accelerator busy to the best extent possible avoids major dark silicon, even if low-reuse sub-accelerator is idle in cases where there are more high-reuse operations.

\textbf{Inter-cascade Partitioning:}
Inter-cascade partitioning has been popularized by prior works on heterogeneous dataflow accelerators\cite{kwon2021heterogeneous,qin2022enabling} in the context of having heterogeneity in terms of different dataflows used.  Within transformers themselves, HHP benefits from inter-cascade reuse as ~\autoref{fig:partitioning}(b) shows. Prefill and decode stages are mapped on high- and low-reuse sub-accelerators respectively. Unlike BERT, L and A operations in the prefill stage, are mapped on high-reuse accelerator here since the decode stage has 1-2 orders of magnitude lower reuse. \autoref{fig:partitioning} shows this difference on a roofline.

\insertWideFigure{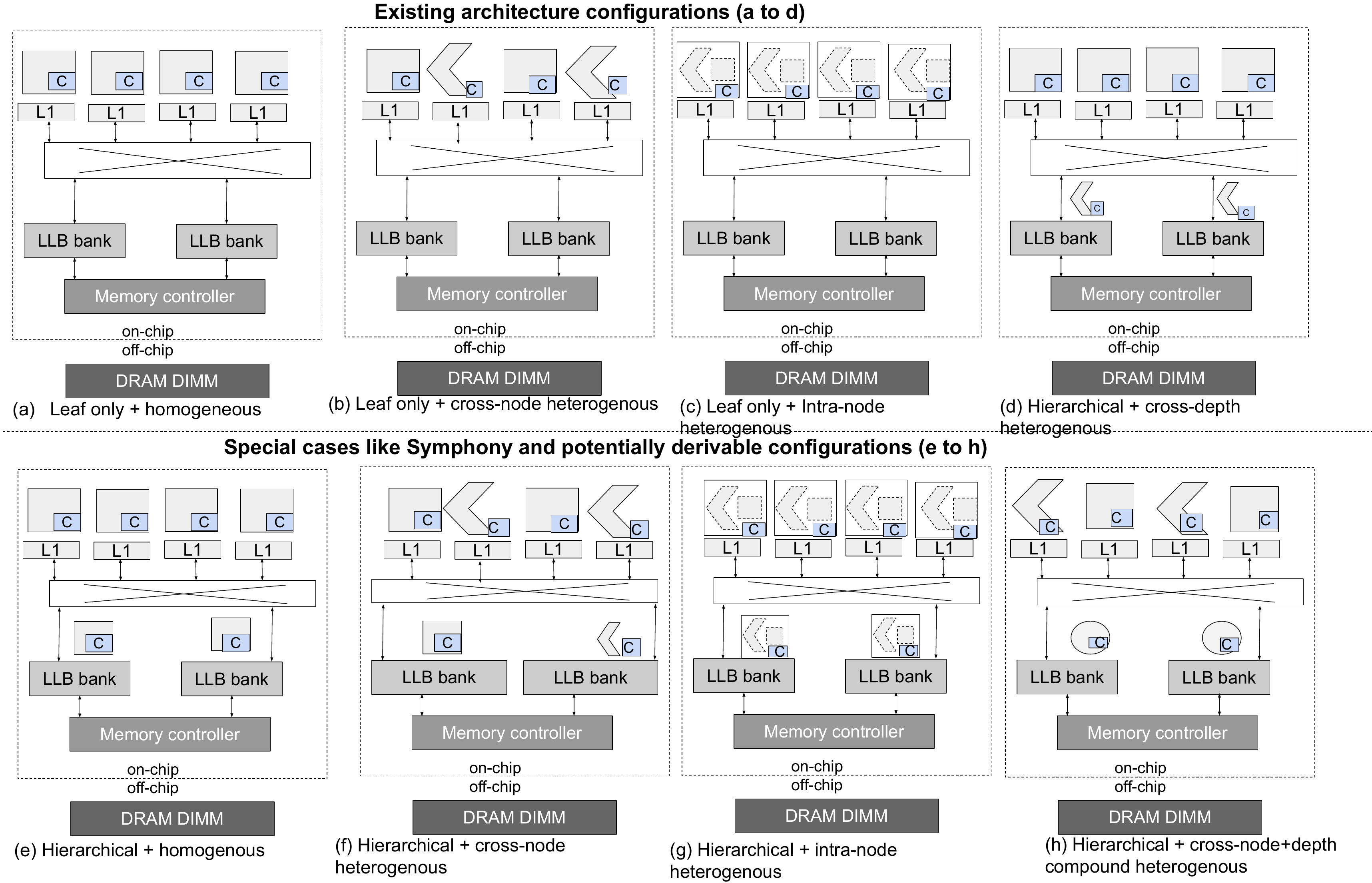}{Various examples hierarchical and/or heterogeneous processors described by \HHPName taxonomy. The square and chevron shapes represent different sub-accelerator architectures, while C represents the FSM controller tied to the sub-accelerators.}

\begin{table*}[t]
\begin{scriptsize}
    \centering
        \caption{Existing works classified by the taxonomy. Some examples from~\autoref{fig:taxonomy} ((e),(g),(h)), are not exhibited by any prior work.}
    \begin{tabular}{|p{0.02\linewidth}|p{0.08\linewidth}|p{0.13\linewidth}|p{0.24\linewidth}|p{0.40\linewidth}|}\hline
         & \textbf{Hierarchical?} & \textbf{Heterogeneity location} & \textbf{Works} & \textbf{Remarks} \\\hline
     (a) &     Leaf-only & Homogeneous & TPUv1~\cite{tpuv4}, MAERI~\cite{kwon2018maeri}, Eyeriss~\cite{eyeriss2016isca}, Flexagon~\cite{flexagon} & Most of the works on DNN accelerators, specially the earlier works. \\\hline

(b) & Leaf-only & Cross-node & Herald~\cite{kwon2021heterogeneous}, AESPA~\cite{qin2022enabling}, TPUv4~\cite{tpuv4} & Herald and AESPA are accelerators with different sub-accelerators good at processing different operation shapes. TPUv4 consists of a sparse and a dense sub-accelerator.  \\\hline
          
  (c) &    Leaf-only  & Intra-node & NVIDIA B100 GPUs~\cite{blackwell}, VEGETA~\cite{vegeta}, RaPiD~\cite{rapid} & NVIDIA GPUs with tensor cores have two different sub-accelerators under a common FSM with the same program counter, thus each node itself is heterogeneous. \\\hline
        
   (d) &     Hierarchical & Cross-depth & NeuPIM~\cite{neupim}, Duplex~\cite{duplex} & NeuPIM and Duplex are hierarchical and heterogeneous, with NPU's at the leaf, and processing-in-DRAM at the root\\\hline

(e) & Hierarchical & Homogeneous & N/A & No prior work fits this category\\\hline
   
   (f) &    Hierarchical & Cross-node & Symphony~\cite{symphony} & Symphony is cross-node heterogeneous, like Herald and AESPA, but still its different since symphony has clustered cross-node heterogeneity. This means that even through, different sub-accelerators are across different nodes, they still are heterogeneous within a cluster that spans a part of the level and repeats itself across the level. \\\hline
   (g)   &  Hierarchical & Intra-node & N/A & No prior work fits this category\\\hline

     (h) &    Hierarchical/leaf-only & Compound heterogeneous & N/A & No prior work fits this category\\\hline
    \end{tabular}
    \label{tables:notation}
\end{scriptsize}

\end{table*}

\section{\HHPName Taxonomy}
\label{sec:taxonomy}

\autoref{fig:taxonomy} shows the proposed \HHPName taxonomy for classifying heterogeneous and/or hierarchical processors. The building blocks (sub-accelerators) in the figure are represented by square, chevron and circle, with change in shape representing heterogeneity. The box with the C represents the position of the FSM controller.

\subsection{Classification of HHP's}

We categorize them on the basis of two axes.

(1) \textbf{Leaf-only}\footnote{Many terms in the taxonomy are derived from treating memory hierarchy as a tree-like structure with DRAM being at the root and L1 being at the leaves with L2 at the intermediate level.}\textbf{ vs hierarchical:
}

\textit{Leaf-only:} Leaf-only implies traditional architectures with computation only at leaves of the memory hierarchy, i.e. closest to the L1 buffer.  Most of the accelerators fall into this category and some of the examples include TPUv1~\cite{tpu-isca}, TPUv4~\cite{tpuv4}, Herald~\cite{kwon2021heterogeneous}, NVIDIA B100 GPU's~\cite{blackwell} etc. \autoref{fig:taxonomy}a)-c) represent leaf-only accelerators where compute is close to only L1.

\textit{Hierarchical:} Hierarchical implies that the computation is distributed across multiple levels of the memory hierarchy. Heterogeneous architectures with processing-in-memory (NeuPIM~\cite{neupim}, Duplex~\cite{duplex}, etc.) are a major example of this. Another example, of hierarchical architectures is Symphony~\cite{symphony} which has "logical elements" for compute across the levels of the memory hierarchy. ~\autoref{fig:taxonomy}d)-h) show examples of hierarchical accelerators with sub-accelerators across memory hierarchy.

(2) \textbf{Location (or absence}\footnote{in case of homogeneous architectures}\textbf{) of heterogeneity:} 

\textit{Homogeneous: }One of the subcategories of this is homogeneous accelerator, where heterogeneity is absent. An example of this is TPUv1~\cite{tpu-isca}. The rest of the categories are based on the source of heterogeneity in the overall structure. \autoref{fig:taxonomy}a) and e) show examples of homogeneity.

\textit{Intra-node heterogeneous}: The finest-grained source of heterogeneity is intra-node heterogeneity(with tree-like treatment of memory hierarchy). In this category, sub-accelerators share a common FSM. An example of this from real-systems is NVIDIA B-100 where the tensor-core operates on the same program counter as the Streaming multiprocessor (SM). Another example of intra-node heterogeneity is a RaPiD core~\cite{rapid} with a 2-D array of MAC units along with a 1-D array to do specialized vector operations that require higher bit precision. \autoref{fig:taxonomy}c) and g) show examples of intra-node heterogeneity.

\textit{Cross-node heterogeneous}: Heterogeneity can also occur across nodes of a tree as opposed to within node. This is the most common form of heterogeneity among the architectures that are not homogeneous, and an example of this in the traditional sense is a CPU-GPU heterogeneous system. Among accelerators, examples include Symphony~\cite{symphony}, Herald~\cite{kwon2021heterogeneous}, AESPA~\cite{qin2022enabling}, TPUv4~\cite{tpuv4}, etc. \autoref{fig:taxonomy}b) and f) show examples of cross-node heterogeneity.

\textit{Cross-depth heterogeneous}: The coarsest form of heterogeneity occurs across different levels of memory hierarchy. An example of this is NeuPIM~\cite{neupim} which consists of an NPU at the leaves for GEMMs and uses processing-in-memory (DRAM) for vectors. \autoref{fig:taxonomy}d) shows a different example of this with sub-accelerators at L1 and L2.

\textit{Compound heterogeneous:} The above sources of heterogeneity are not mutually exclusive, and it is possible to use this framework to derive a new kind of architeture, with multiple sources of heterogeneity. \autoref{fig:taxonomy}h) shows an example of a combination of cross-node and cross-depth heterogeneity, where leaves have different sub-accelerators, and both of these are different from the accelerator at L2. Another example of compound heterogeneous accelerator could include a combination of cross-depth and intra-node heterogeneity with just one of the sub-accelerators having intra-node heterogeneity.

\textbf{Example datapoints:} Architectures are described using the two axes discussed above. For example, leaf-only + cross-node heterogeneous architectures, as the name suggests, exhibit cross-node heterogeneity and only have compute at the leaves as~\autoref{fig:taxonomy}b) shows with squares and chevrons at different nodes among the leaves. Another example is hierarchical+cross-depth heterogeneous, as~\autoref{fig:taxonomy}d) shows, where chevrons and squares are at different levels of the memory hierarchy. Note, that cross-depth heterogeneous is the only category that cannot have a leaf-only counterpart, since compute at atleast two levels is needed for cross-depth heterogeneity. \autoref{fig:taxonomy}c) shows leaf-only+intra-node heterogeneous architecture, where heterogeneity is within a node. Here, square and chevron share a common FSM. An example of this is a GPU with a tensor core. In~\autoref{sec:eval}, we mainly focus on configurations in~\autoref{fig:taxonomy} (a-d), since they cover all the axes, while showing through ~\autoref{fig:taxonomy} (e-h) that we can classify more complex architectures like Symphony~\cite{symphony} and derive new categories using the taxonomy.

\subsection{Describing Existing Works}

\autoref{tables:notation} shows some examples of existing architectures, described using the \HHPName taxonomy. The table only describes existing work and does not cover some of the combinations from~\autoref{fig:taxonomy}. These combinations are plausible, but have not been exhibited in prior works. Most of the earlier accelerators fall under the leaf-only+homogeneous category. These works include simpler accelerators with fixed-dataflow like TPUv1~\cite{tpu-isca} and Eyeriss~\cite{eyeriss2016isca}, and flexible accelerators that support multiple dataflows (e.g. Flexagon~\cite{flexagon} and MAERI~\cite{kwon2018maeri}). The next most common category is leaf-only+heterogeneous accelerators. Herald~\cite{kwon2021heterogeneous} was one of the earlier works to propose a heterogeneous accelerator with each sub-accelerator being good for different CONV operation shapes. AESPA~\cite{qin2022enabling} proposed a similar cross-node heterogeneous accelerator but for SpGEMMs. These works can be described as leaf-only+cross-node heterogeneous accelerators. Leaf only+ Intra-node heterogeneous accelerators are also popular. These works typically consist of the sub-accelerators sharing a common FSM. For example, in NVIDIA B100, SM's and tensor-core share an FSM. This is the most tightly coupled form of heterogeneity. Hierarchical accelerators have recently emerged in popularity. NeuPIM~\cite{neupim} and Duplex~\cite{duplex} directly target transformers with mixed-reuse operations and exhibit cross-depth heterogeneity. Symphony~\cite{symphony} is hierarchical and exhibits cross-node heterogeneity. However, even though, the sub-accelerators have different FSM's, unlike herald, sub-accelerators are cross-node within a cluster, and that cluster is then repeated across the level. Herald on the other hand consists of entire sub-accelerator in one local area.

\section{Deeper Dive: Analysis of HHP's}


The \HHPName taxonomy provides a structured way of classifying the architectures based on the relative location of sub-accelerators and levels of memory hierarchy that have compute.

\subsection{Pros and Cons of Heterogeneous Architectures}

Heterogeneous accelerators are good at decoupling high-reuse and low-reuse accelerators and they make it easier for the mappers to assign high-reuse task to one unit and low-reuse task to another unit. However, for an area normalized setup, both individual sub-accelerators have lower compute roof compared to a homogeneous accelerator with same number of resources. 

If the workload is sufficiently compute-bound or if the memory bandwidth is sufficiently high, homogeneous accelerator utilizes its complete compute roof to achieve undivided throughput. However, the heterogeneous setup is beneficial when the compute-roof gap between the operations is as significant as 1-2 orders of magnitude. In that case, the homogeneous accelerator wastes its compute roof, while a perfect overlap between heterogeneous sub-accelerators leads to better utilization.

Another factor that affects the performance of heterogeneous accelerator is the dependency graph. For example, BERT only has a few opportunities to overlap two operations in an intra-cascade fashion (e.g., value matrix generation and logit operations). Moreover, typically the maximum sequence length is less than the hidden dimension of the model ($L_{max}<d_{model}$), which implies that the compute volume of one BMM operation is lower than one GEMM operation. Moreover, the number of GEMM operations in an attention layer itself is twice the number of BMM operations. This exacerbates the compute volume gap beween high and low-reuse operations in BERT. Thus, we observe that, with sufficiently high memory bandwidth, heterogeneous accelerators are less effective on workloads like BERT.
GPT on the other hand, has pre-fill and decode stages that can be overlapped in an inter-cascade fashion. Decode stage has lower reuse than prefill stage. These sub-cascades can be executed independently, and take advantage of the overlap that heterogeneous sub-accelerators offer, thus minimizing the wastage of compute roof in homogeneous accelerators.

\subsection{Impact of heterogeneity position}

The biggest advantage of hierarchical architecture is ease of decoupling high- and low-reuse sub-accelerators. This is especially useful when high- and low-reuse operations have different shapes. Intra-node architecture on the other extreme has sub-accelerators coupled with most of the parallelism strategies being common due to the shape. For example, in a RaPiD-like~\cite{rapid} setup, the only parallelism dimension that is different is the number of rows in the PE array and the Special Function Unit array. This easily works for element-wise operations followed by DNN layers, since they use the input generated by the previous operation, preserving the problem dimensions across sub-accelerators. However, repurposing it for two different operations with different reuse strategies poses mapping challenges, as we also see in one of the results in~\autoref{sec:eval}. Cross-depth heterogeneous accelerators have the lowest energy, since they avoid data movement across an entire level of memory hierarchy.

\insertWideFigureScaled{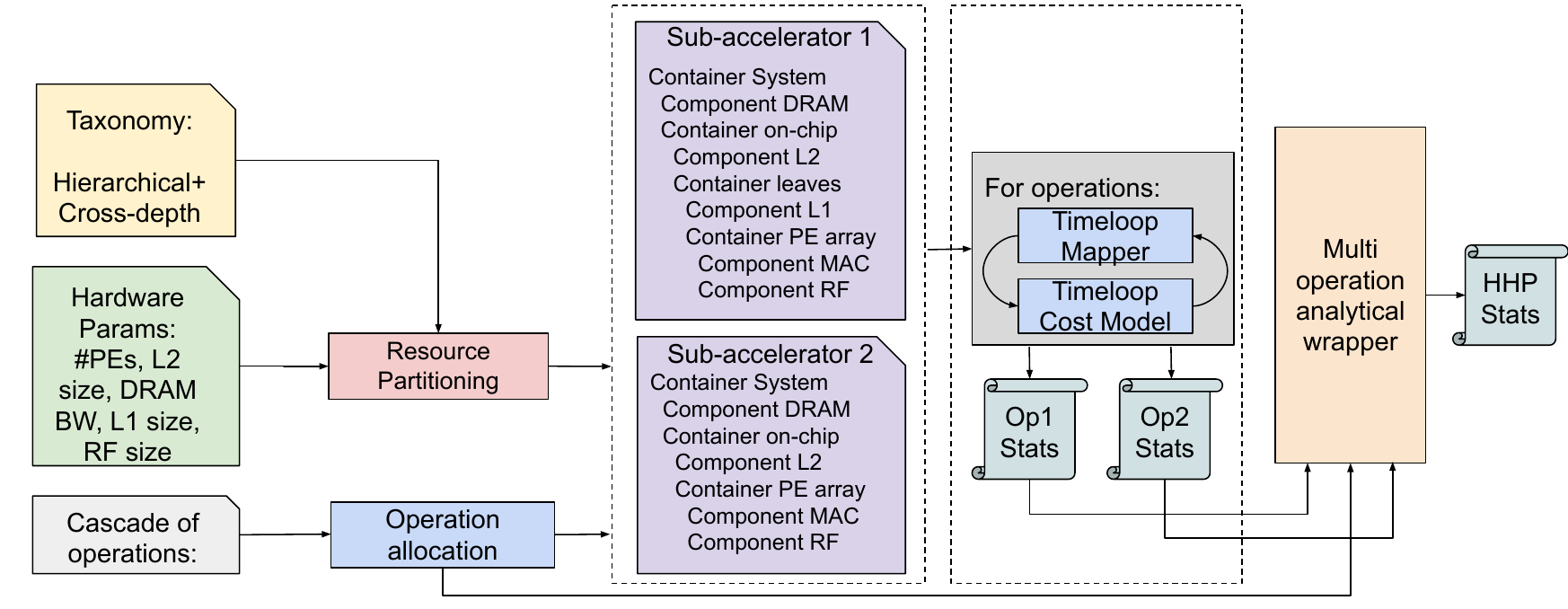}{Evaluation framework built on Timeloop~\cite{timeloop}.
}{0.85}


\subsection{Mapping on HHP's}
\label{sec:mapping}

In this work, while we run the timeloop mappers independently on sub-accelerators, we make use of spatial parallelism mapping constraints to capture relative position of heterogeneity. For example, in a RaPiD~\cite{rapid}-like\footnote{Note in actual RaPiD accelerator, the functional units are used for element-wise and high precision operations, but the concept of two sub-accelerators in one node can be used to compute mixed-reuse cascades.} intra-node accelerator, the number of columns per sub-accelerator are equal, and the same dimension can be parallelized across columns. Cross-depth accelerator on the other hand, allows for complete decoupling of spatial parallelism strategies.

Since we partition the workload in an operation-by-operation manner and assign it to sub-accelerators based on reuse, the operation mapping search can be carried out independently on sub-accelerators, which we refer to as \textit{blackbox} mapping. Therefore the design-space is additive and not multiplicative. For example, in case of one high-reuse operation and one low-reuse operation, the design-space is $O(High+Low)$ and not $O(High\times Low)$

\subsection{Resource Partitioning}

For heterogeneous accelerators, our framework (\autoref{sec:expt}) partitions the compute roof (number of PEs), L1 and L2 resources, and memory bandwidth. We observe that the performance of heterogeneous accelerators is highly sensitive to memory bandwidth. For cascade(s) where the latency of low-reuse operations dominate (e.g. GPT3), low-reuse sub-accelerators should get a larger portion of the memory bandwidth. However, if the number of high-reuse operations is larger (e.g. BERT), allocating higher memory bandwidth to low-reuse operations, can affect the performance of high-reuse operations which primarily determine the latency. In such cases, the bandwidth partitioning is governed by two conflicting forces - (1) high-reuse operations dominate the cascade, and (2) low-reuse operations are in need of higher portion of the bandwidth, to improve the operation latency. Across the memory hierarchy, we partition the LLB in the ratio of compute roof, as high-reuse operations benefit from more on-chip memory space while low-reuse operations peak their arithmetic intensity even with low on-chip reuse space. In case of hierarchical accelerator, L1 is used purely by the high-reuse sub-accelerator, and is not partitioned.

\section{Experimental Methodology}

\label{sec:expt}

\subsection{Evaluation Framework: Timeloop for \HHPName}

\autoref{fig:framework} shows the evaluation framework built on Timeloop~\cite{timeloop}. The inputs to the framework include configuration based on the proposed \HHPName taxonomy, cascades of operations, and hardware parameters. Operations are allocated based on reuse to sub-accelerators, and based on the taxonomy notation and resource partitioning, sub-accelerator architecture files are generated. These sub-accelerator files are based on Timeloop v0.4 and include a detailed hardware model with the ability to specify mapping constraints and resources available to each sub-accelerator. Timeloop mapper is called for each operation with the architecture files corresponding sub-accelerators executing that operation. We use a wrapper to compute the statistics of the HHP configuration from statistics of operations executed on individual sub-accelerators to return the final results corresponding to the cascade.

\subsection{Workloads}

We evaluate HHP's on mixed-intensity cascades of operations as shown in~\autoref{tables:workloads}. We focus on transformers, where we consider an attention layer from encoder-only transformers as intra-cascade workload partition, and prefill and decode stages from decoder-only transformer as inter-cascade workload partition. 

\begin{table}[h]
\begin{scriptsize}
    \centering
    \caption{Transformer workload configuration. Sequence lengths for decoder-only transformers are listed as prefill/decode}
    \begin{tabular}{|c|c|c|c|c|}
    \hline
         \textbf{Workload} & \textbf{Models} & \textbf{Partitioning} & {$\mathbf{d_{model}}$} &
         \textbf{Seq length} \\\hline
         Encoder (translation) & BERT-large & Intra-cascade & 1024 & 256\\\hline
         Decoder (chatbot~\cite{genz}) & Llama-2 & Inter-cascade & 4096 & 3000/1000\\\hline
         Decoder (chatbot~\cite{genz}) & GPT3  & Inter-cascade & 12288 & 3000/1000\\\hline
          
    \end{tabular}
    \label{tables:workloads}
\end{scriptsize}
\end{table}

\subsection{Evaluation Configurations}

We evaluate the common configurations configurations based on the taxonomy shown in~\autoref{fig:taxonomy} (a-d) to study leaf vs hierarchical and homogeneous vs cross-node heterogeneous vs intra-node heterogeneous vs cross-depth heterogeneous. The hardware parameters are described in~\autoref{tables:config}. Other parameter changes for specific sensitivity studies are mentioned along with the studies.

\begin{table}
\begin{scriptsize}
        \centering
    \caption {Hardware parameters.}
    \begin{tabular}{|c|c|}
    \hline
    \textbf{Parameter} & \textbf{Value}\\\hline
       Datawidth (bits per word) & 8\\\hline
       Number of MACs  &  40960\\\hline
       Read DRAM bandwidth (bits per cycle) & Sweep: 2048, 512  \\\hline
       Write DRAM bandwidth (bits per cycle) & Sweep: 2048, 512  \\\hline
       R/W DRAM bandwidth (bits per cycle) & Sweep: 2048, 512  \\\hline
       Shared DRAM bandwidth (bits per cycle) & Sweep: 2048, 512  \\\hline
       LLB size  & 4MB \\\hline
       L1 size (per array)  & 0.125 MB\\\hline
       RF size (per PE)  & 64B\\\hline
       High:Low reuse compute roof ratio & 4:1\\\hline
    \end{tabular}
    \label{tables:config}
    \end{scriptsize}
\end{table}

\insertWideFigureScaled{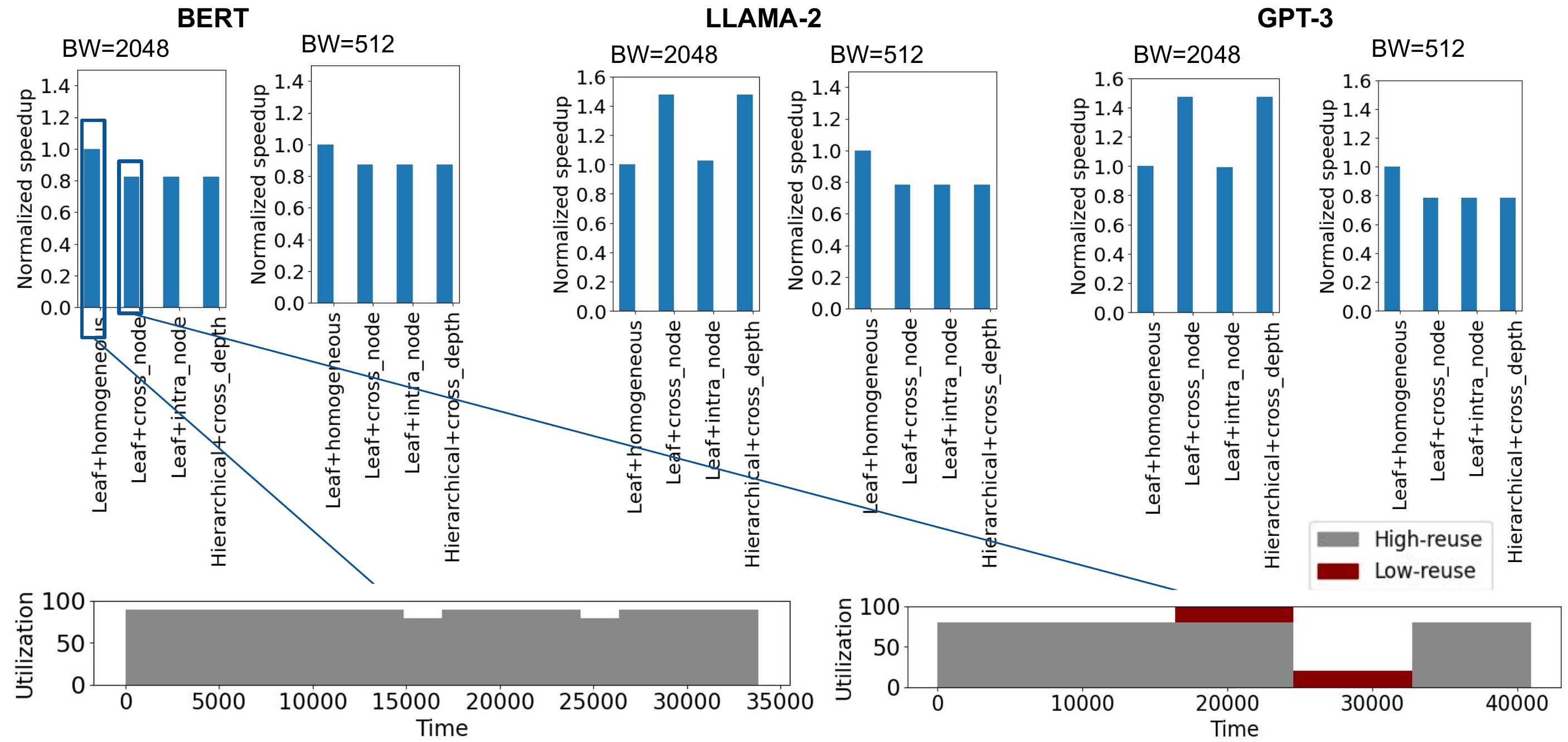}{Speedup of various architecture configurations normalized to leaf+homogeneous. We zoom in on the utilization of homogeneous and cross-node heterogeneous accelerators on BERT.}{1}

\insertFigureScaled{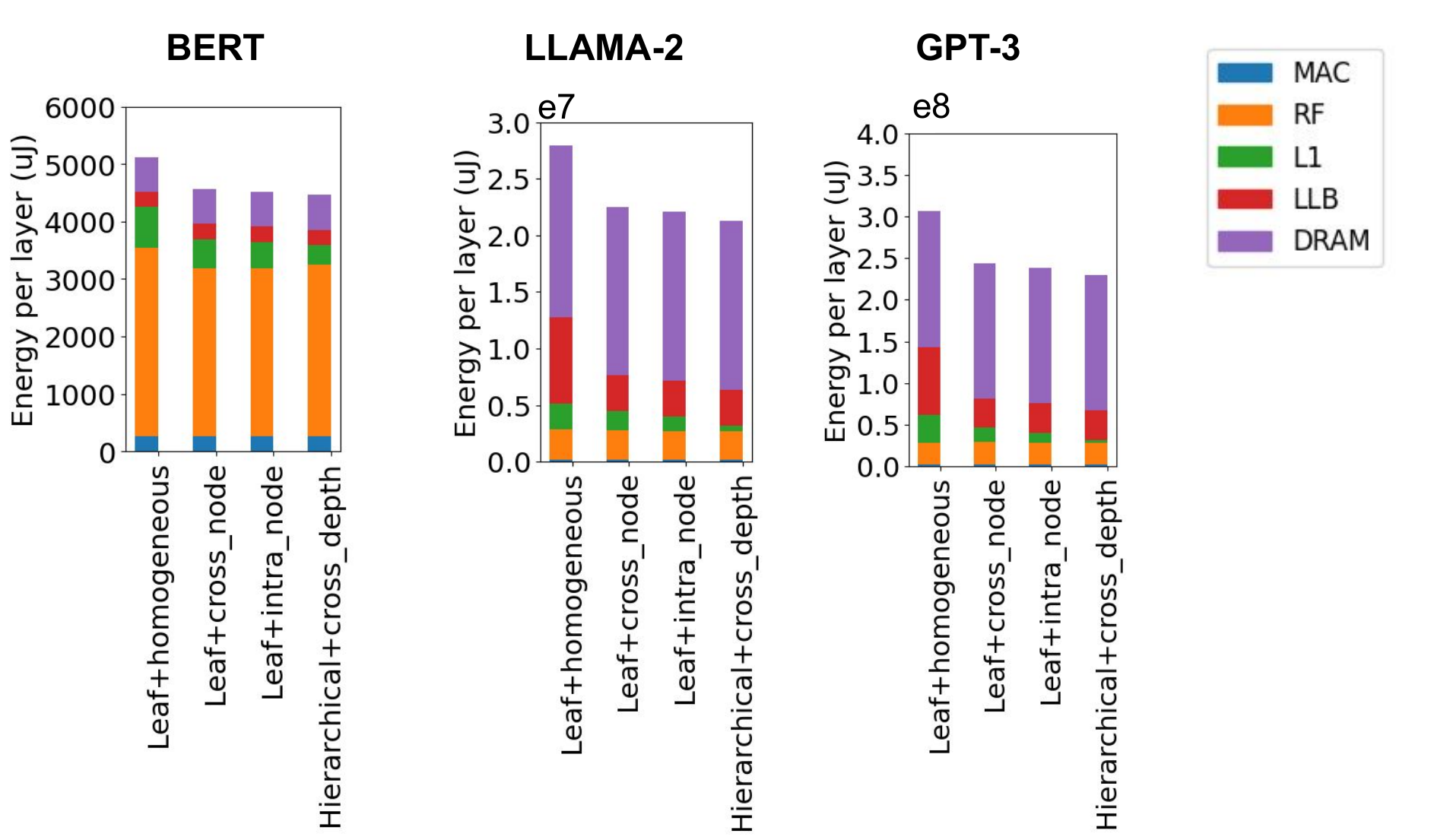}{Energy of various architecture configurations.}{0.99}

\insertFigureScaled{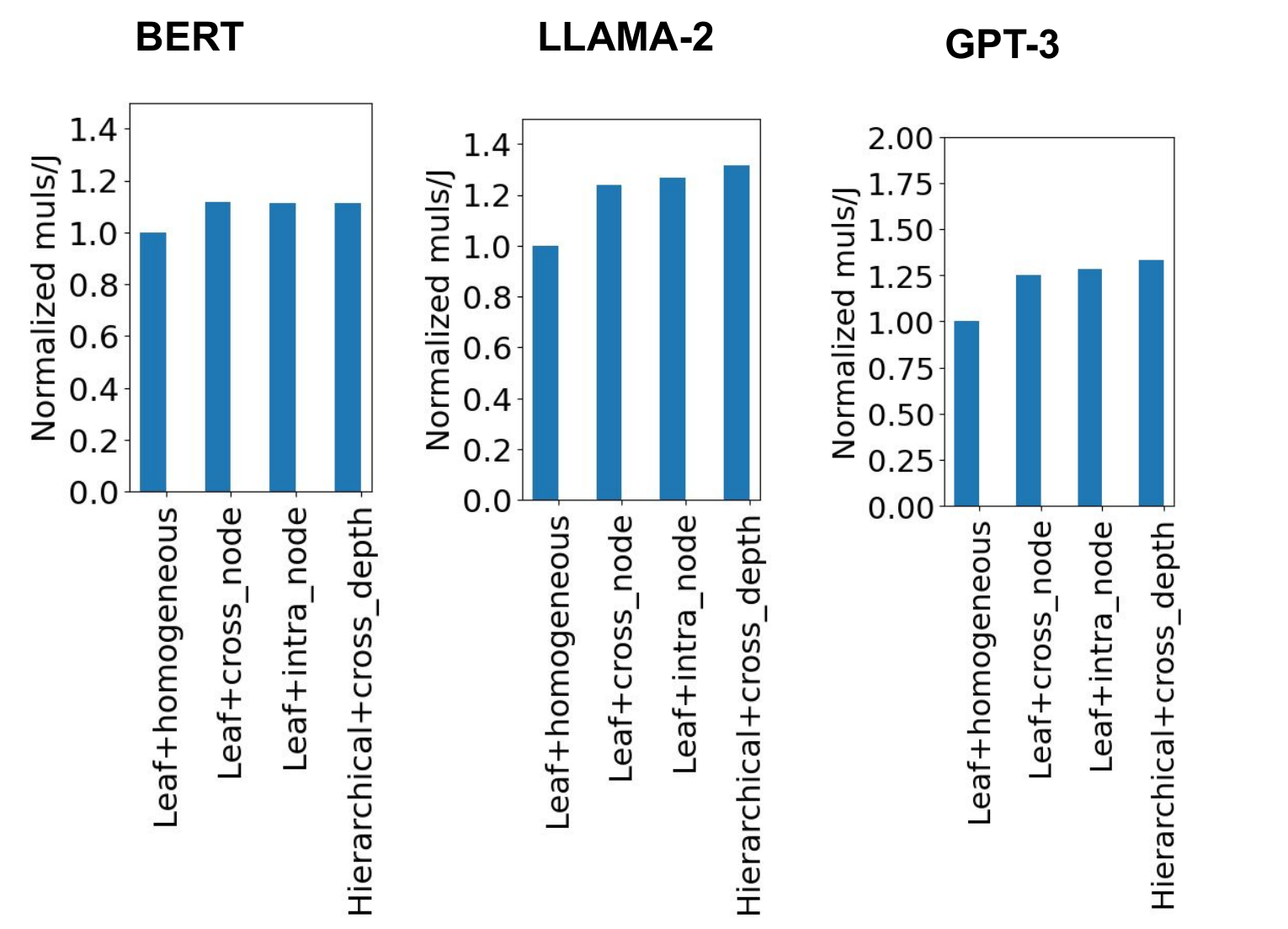}{Energy per joule for various architecture configurations normalized to leaf+homogeneous}{0.99}

\insertFigureScaled{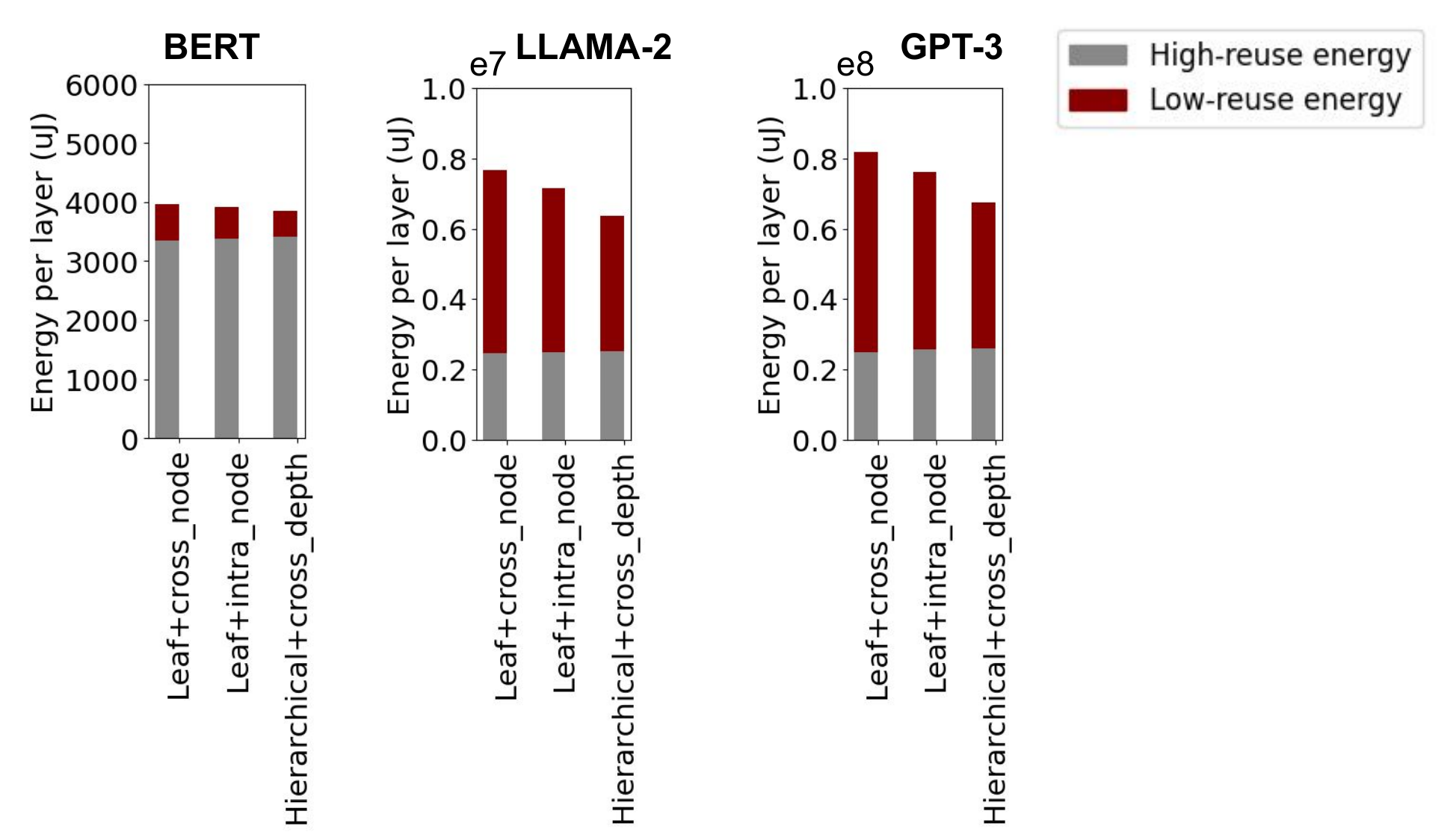}{On-chip energy (excluding DRAM) breakdown by sub-accelerators running high-reuse and low-reuse operations.}{1}

\insertFigureScaled{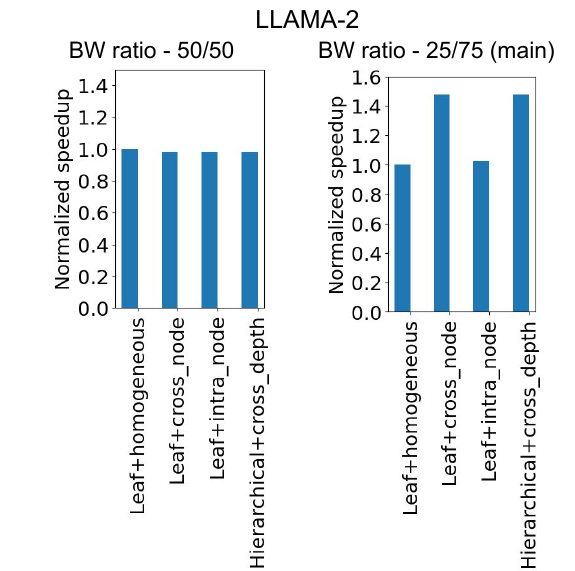}{Impact of 50/50 bandwidth partitioning in case of decoder-only transformer.}{0.8}

\section{Results}
\label{sec:eval}

\subsection{Performance}

~\autoref{fig:eval-perf} shows performance of HHP's for transformer models. The performance of each workload depends on multiple factors, and there is no one size fits all architecture for all applications. Under the normal bandwidth of 2048 bits/cycle, encoder-only workload performs better on homogeneous configuration compared to heterogeneous counterpart since the number of high-reuse operations is larger, and the overlap potential is lower. We zoom into the utilization over time by homogeneous and cross-node accelerators on BERT, and observe that the ability to overlap high-reuse and low-reuse accelerator can affect the performance. We also find that the relative speedup of homogeneous accelerator deteriorates with lower bandwidth. We find that this is because of lower utilization of the PEs when the low-reuse workload is run. However, in case of decoder only workloads, the performance trends are reversed. This is because of better decoupling of high- and low-reuse subcascades in the form of prefill and decode stages. Heterogeneous sub-accelerators are able to utilize the PEs in a better way by parallelizing high- and low-reuse cascades. The low-reuse sub-cascade is slower than high-reuse sub-cascade. However, we also observe that tight coupling of sub-accelerators in intra-node configuration leads to lower utilization in case of decoder-only workload. On lowering the memory bandwidth, we find that the low-reuse sub-accelerator of the heterogeneous configuration gets hit harder, since it is using only a portion of the memory bandwidth in parallel with the high-reuse sub-accelerator. This trend is completely opposite of the impact of lowering bandwidth on encoder-only workload, since in case of encoder-only workload, low-reuse accelerator was already running at lower compute roof with 100\% utilization of that sub-accelerator and was not affected by lowering the bandwidth. In case of the decoder-only model, the achieved utilization was 19\% of the compute-roof and was further hit after lowering the bandwidth by 4x. 

\subsection{Energy}

~\autoref{fig:eval-energy} shows the energy of various architectures broken down across levels of memory hierarchy. The heterogeneous accelerator configuration has lower overall energy consumption than the homogeneous accelerator configuration for encoder and decoder workloads by roughly 10\% and 20\% respectively. For encoder-only models, the energy is dominated by register-file (RF) which shows good overall reuse, while for decoder-only models, the energy is significantly dominated by DRAM accesses, showing less local reuse opportunities. Hierarchical+cross-depth architecture has the least energy since the low-reuse sub-accelerator avoids data movement at one level of the memory hierarchy. For encoder-only cascades, the difference in the energy mainly stem from the register file energies, while in case of decoder-only cascades, the major parameters the differences stem primarily from LLB energies, and L1 energy in case of hierarchical+cross-depth heterogeneous accelerator.

\subsection{Multiplications Per Joule}

~\autoref{fig:eval-efficiency} shows multiplications per joule (energy efficiency) of various accelerator configurations. Cross-depth heterogeneous accelerator is the most energy efficient architecture, while the homogeneous architecture (with same number of total PEs as heterogeneous sub-accelerator) is the least energy efficient. The energy efficiency difference is more pronounced in case of GPT, where low-reuse operations dominate the energy.

\subsection{Energy Breakdown Across Sub-accelerators}

~\autoref{fig:eval-energy2} shows the on-chip energy breakdown across sub-accelerators in case of intra-node, cross-node and cross-depth heterogeneity. For encoder-only model (BERT), we observe that high-reuse operations contribute significantly to the energy, while for decoder-only models (GPT3 and Llama-2), we observe that low-reuse operations contribute more significantly to the energy.

\subsection{Sensitivity Study - Bandwidth Partitioning}

We also compare the bandwidth partitioning in case of decoder-only workload. We originally allocated 75\% memory bandwidth to low-reuse sub-accelerator in case of decoder-only workload because decode stage is significantly slower. However, allocating 50\% bandwidth to each sub-accelerator naively erodes the advantage heterogeneous configuration had by overlapping high- and low-reuse operations as~\autoref{fig:eval-bw} shows.

\subsection{Summary of Key Trends}

The key observations can be summarized as follows:

\squishlist
\item For encoder-only model (BERT), homogeneous accelerator has better performance since intra-cascade partition can depend on the dependency graph. For decoder-only models (GPT and Llama), heterogeneous accelerator performs better because of the ability to overlap high- and low-reuse operations. Decoder-only models significantly dominate latency.
\item Hierarchical+Cross-depth accelerators have the lowest energy and the highest energy efficiency.
\item The energy is dominated by DRAM in case of decoder models and register file in case of encoder models
\item Heterogeneous accelerators are sensitive to memory bandwidth partitioning.
\item On-chip energy is dominated by high-reuse operations for applications like BERT, while it is dominated by low-reuse operations for applications like Llama.
\squishend
\section{Additional Related Work}

\autoref{tables:notation} shows prior works on hierarchical and heterogeneous accelerators. Other works like MAESTRO~\cite{kwon2019understanding}, Eyeriss~\cite{eyeriss2016isca} and OMEGA~\cite{garg2021understanding} have proposed taxonomies to classify scheduling strategies. Herald~\cite{kwon2021heterogeneous} proposes a heterogeneous accelerator framework based on cross-node-only leaf, but partitions the workload based on optimal spatial parallelism strategies. Works like Sparseloop~\cite{sparseloop}, 
LayoutLoop~\cite{feather}
CimLoop~\cite{cimloop} have also modified Timeloop to enable exploration in new domains like sparsity, data layout, compute-in-memory and so on.

\section{Conclusion}
\label{sec:discussion}

Mixed-reuse applications are an emerging trend in the field of AI, with transformers being the most popular. Hierarchical and heterogeneous architectures have is an emerging way to run operations of different arithmetic intensities. In this work, we propose a taxonomy to classify hierarchical and heterogeneous processors (HHP's) based on the position of compute across different levels of memory hierarchy and on the location on (or absence of) heterogeneity. We modify timeloop to enable blackbox mapping on sub-accelerators in a hierarchical or heterogeneous and study the performance and energy of these architectures for transformer workloads. We find that the performance and energy trends across HHP's greatly vary depending on the type of transformer (encoder-only or decoder-only), memory bandwidth partitioning and even the energy breakdown varies across workloads and architectures.  We envision that the taxonomy and proposed framework will enable exploration of new accelerators for mixed-reuse applications.

\section*{Acknowledgment}
This work was supported by ACE, one of the seven centers in JUMP 2.0, a
Semiconductor Research Corporation (SRC) program sponsored by DARPA.
\clearpage
\bibliographystyle{IEEETranS} 
\bibliography{refs}


\end{document}